\documentclass[11pt]{article}
\usepackage[margin=1in]{geometry}
\usepackage{amsmath,amssymb,amsfonts}
\usepackage{graphicx}
\usepackage{booktabs}
\usepackage{siunitx}
\usepackage[colorlinks=true,linkcolor=blue,citecolor=blue,urlcolor=blue]{hyperref}
\usepackage[numbers,sort&compress]{natbib}
\usepackage{setspace}
\usepackage{authblk}
\usepackage{caption}
\usepackage{bm}
\usepackage{array}
\usepackage{mathtools}
\graphicspath{{figs/}}
\title{\bfseries Relativistic PBTE:\\
Biological Proper Time Along the Worldline}
\author[1]{Mesfin Asfaw Taye\thanks{Correspondence: \texttt{tayem@wlac.edu}}}
\affil[1]{Science Division, West Los Angeles College, 9000 Overland Ave, Culver City, CA 90230, USA}
\date{\today}

\begin{document}
\maketitle

\begin{abstract}
\noindent
Biological aging is conventionally indexed by chronological time, yet every organism is a physical
system tracing a worldline through spacetime, so the time available to its metabolism is not
coordinate time $t$ but relativistic proper time $\tau$. Building on the established Principle of
Biological Time Equivalence (PBTE), whose thermodynamic foundation and aging dynamics are taken
here as prior results \citep{TayeBook2026,TayeAging2026,TayeCardiac2026}, we ask how internal
physiological time relates to the proper time of physics. The central result is that biological
age is an entropy-production functional evaluated along the proper-time worldline,
\begin{equation*}
A_{\rm PBTE}=\frac{1}{\Sigma_{\rm ref}}\int_{\tau_0}^{\tau_1}\dot\Sigma_p(\tau)\,d\tau,
\qquad
\frac{dA_{\rm PBTE}}{dt}=\frac{\dot\Sigma_p}{\gamma\,\Sigma_{\rm ref}},
\end{equation*}
where $\dot\Sigma_p$ is entropy production per unit proper time in the local rest frame and
$\gamma$ the Lorentz factor. The consequence is that relativistic time dilation is not caused by
physiology: physiology, like every physical process, is parametrized by proper time---a fact
established on muons, atomic and optical clocks, and GPS, none of them living
\citep{Bailey1977,HafeleKeating1972a,Chou2010,Ashby2003}---and PBTE only fixes the irreversible
biological history accumulated per unit of it. The word \emph{relativity} is used here in two
compatible senses: the standard relativity of proper time, supplied by spacetime geometry, and a
distinct \emph{PBTE relativity} that is relational rather than geometric---two organisms exposed
to the same external interval need not accumulate the same biological interval, because their
characteristic physiological frequencies, entropy costs per cycle, and effective lifetime budgets
differ, with internal times transformed by $d\theta_i/d\theta_j=f_i/f_j$. The framework preserves
Einstein relativity intact, separates physiological from geometric contributions to externally
observed aging, and yields falsifiable predictions for organisms in relativistic motion or
differing gravitational potentials. More broadly, it reframes the comparative question of biology
from how many years an organism has existed to how much internally generated biological history
it has traversed.
\end{abstract}
\section{Introduction}
\label{sec:introduction}

Time enters modern science through several logically distinct concepts.
In Newtonian mechanics it is an absolute external parameter, assumed to
advance uniformly for every observer \citep{Newton1687}. Special and
general relativity replace this universal background with proper time:
the invariant duration accumulated by a physical system along its
worldline through spacetime \citep{Einstein1905,Einstein1916,MTW1973,
Rindler2006}. Thermodynamics introduces a different structure, the arrow
of time, through the macroscopic irreversibility associated with entropy
production \citep{Eddington1928,Prigogine1961,Seifert2012}. Biology adds
another level. Living systems possess internal rhythms---cardiac cycles,
respiration, circadian oscillations, cellular division, and molecular
turnover---whose cumulative progression need not remain proportional to
calendar time across organisms or physiological states
\citep{Lindstedt1981,Calder1984,SchmidtNielsen1984}.

The Principle of Biological Time Equivalence (PBTE) was introduced to
formalize this distinction. Its empirical point of departure is the
approximate compensation between physiological frequency and lifespan.
A mouse lives for only a few years and possesses a rapid heart rate,
whereas an elephant lives for several decades and possesses a much slower
heart rate. Although their chronological lifespans differ by more than an
order of magnitude, their lifetime cardiac counts are both of order
\(10^{9}\) cycles \citep{Levine1997,Cook2006}. PBTE interprets this
regularity by defining biological proper time as the accumulated number of
intrinsic physiological events and by treating aging as the irreversible
thermodynamic expenditure associated with those events
\citep{TayeBook2026,TayeAging2026,TayeCardiac2026,TayeThermo2026}.

It is useful to fix at the outset the sense in which the present work speaks
of \emph{relativity}, because two compatible meanings are involved.
Relativistic proper time retains its standard physical meaning: the
invariant duration accumulated along a worldline and fixed by spacetime
geometry. PBTE introduces no alternative spacetime, no biological Lorentz
group, and no modification of special or general relativity. Its additional
claim is relational rather than geometric: biological duration is relative
to the internal physiological clock by which a living system progresses
through its own organized history. Two organisms exposed to the same
external interval need not accumulate the same biological interval, because
their characteristic physiological frequencies, entropy costs per cycle, and
effective lifetime budgets may differ. In this restricted but physically
meaningful sense PBTE defines a biological relativity principle, in which the
transformation between organisms is governed not by velocity but by the
ratio of their intrinsic clock rates, \(d\theta_i/d\theta_j=f_i/f_j\), and,
in the entropy-normalized formulation, by the irreversible cost assigned to
those cycles. The mouse--elephant comparison gives this an operational
meaning. If the cardiac frequency of the mouse is approximately twenty times
that of the elephant, then during the same calendar day the mouse
accumulates approximately twenty times as many cardiac cycles; one mouse day
contains about the same cardiac-cycle distance as twenty elephant days. The
two organisms inhabit the same astronomical day but do not spend the same
amount of biological time within it. PBTE therefore shifts the fundamental
comparative question from how many years an organism has existed to how much
internally generated biological distance it has traversed.

The empirical invariant, its thermodynamic closure, and its application to
biological aging have previously been formulated relative to terrestrial
chronological time. That approximation is sufficient for conventional
comparative biology, because relativistic differences between terrestrial
organisms are negligible. It is not, however, a fundamental formulation.
Every organism is also a physical system following a worldline, and every
local physical process within that organism evolves according to the
proper time accumulated along that worldline. A complete formulation of
PBTE must therefore distinguish the external coordinate time \(t\), the
relativistic proper time \(\tau\), the accumulated biological cycle count
\(\theta\), and the entropy-normalized biological age
\(A_{\rm PBTE}\).

This distinction resolves the central causal question addressed in the
present work. A rapidly moving organism does not age more slowly because
uniform velocity directly suppresses its physiology. In the organism's
local rest frame, metabolism, cardiac activity, chemical reactions,
molecular turnover, and cognition proceed normally. The difference
observed from another frame arises because different worldlines contain
different amounts of proper time between specified comparison events.
Relativistic time dilation is therefore prior to physiology rather than
produced by it. This conclusion follows from the experimentally established
behavior of particle lifetimes, transported atomic clocks, optical clocks,
and satellite timing systems, none of which requires a biological
mechanism \citep{Bailey1977,HafeleKeating1972a,HafeleKeating1972b,
Chou2010,Ashby2003}.

The contribution of this paper is to place PBTE consistently within that
proper-time structure. Biological age is defined as an
entropy-production functional evaluated along the organism's worldline.
The resulting formulation separates the geometric and physiological
contributions to externally observed aging:

\begin{equation}
\boxed{
\begin{array}{c}
\text{spacetime geometry determines }d\tau;\quad
\text{local physiology determines }d\Sigma_p;\\[2pt]
\text{PBTE converts }d\Sigma_p\text{ into biological age}.
\end{array}
}
\label{eq:synthesis}
\end{equation}

The central law, derived in Section~\ref{sec:relpbte}, expresses biological
age as an entropy-production functional evaluated along the proper-time
worldline,
\begin{equation}
A_{\rm PBTE}
=
\frac{1}{\Sigma_{\rm ref}}
\int_{\tau_0}^{\tau_1}
\dot{\Sigma}_p(\tau)\,d\tau,
\qquad
\dot{\Sigma}_p \equiv \frac{d\Sigma_p}{d\tau},
\label{eq:central_intro}
\end{equation}
in which \(A_{\rm PBTE}\) is the entropy-normalized biological age,
\(\dot{\Sigma}_p\) the irreversible entropy-production rate per unit proper
time measured in the organism's local rest frame, \(\Sigma_{\rm ref}\) a
reference lifetime entropy budget, and \(\tau\) the relativistic proper
time. We do not modify relativity; we identify the temporal parameter on
which the previously developed PBTE aging functional must be evaluated.
The mathematical operation is elementary, yet its physical content is
consequential: spacetime fixes the proper duration available to the
organism, whereas PBTE fixes the irreversible biological history
accumulated during that duration. The remainder of the paper makes this
precise. Section~\ref{sec:basics} restates the PBTE foundations---the
lifetime-cycle relation, its thermodynamic closure, and the resulting
entropy-normalized age---and isolates the proper-time problem that
motivates the present work. Section~\ref{sec:relpbte} develops the
covariant relativistic functional of Eq.~\eqref{eq:central_intro} and its
physical consequences, and Section~\ref{sec:discussion} states the
resulting temporal hierarchy, the limitations of the framework, and its
falsifiable predictions.

\section{PBTE Foundations and the Relativistic Proper-Time Problem}
\label{sec:basics}

The present study extends the Principle of Biological Time Equivalence
(PBTE) to systems whose motion through spacetime cannot be described by a
single universal time coordinate. It does not propose an alternative theory
of special relativity, nor does it attempt to explain relativistic time
dilation through physiology. Its purpose is more specific. Previous PBTE
work formulated biological progression and thermodynamic expenditure
relative to an external chronological parameter \(t\). Relativity, however,
establishes that the duration locally accumulated by any physical system is
the proper time \(\tau\) along its worldline. The problem addressed here is
therefore one of temporal consistency: how should PBTE be formulated when
coordinate time and proper time are no longer interchangeable?

The empirical foundation of PBTE is the lifetime-cycle relation. Let
\(f_i(t)\) denote the instantaneous frequency of a recurrent physiological
process in organism \(i\). Depending on the biological system under
consideration, this frequency may represent cardiac activity, respiration,
cell division, metabolic cycling, or another recurrent process that provides
an operational measure of physiological progression. The accumulated number
of intrinsic cycles is defined as biological proper time,

\begin{equation}
\theta_i(t)
=
\int_{0}^{t} f_i(s)\,ds,
\qquad
d\theta_i
=
f_i(t)\,dt.
\label{eq:theta_definition}
\end{equation}

The term ``proper time'' is used here in a biological rather than a
spacetime sense. The quantity \(\theta_i\) is dimensionless and counts
intrinsic biological events. It is not a coordinate of spacetime and should
not be identified with the relativistic proper time \(\tau_i\). The
distinction between these two quantities is central to the development
below.

If \(L_i\) denotes the chronological lifespan of the organism, its total
lifetime biological time is

\begin{equation}
N_i
\equiv
\theta_i(L_i)
=
\int_{0}^{L_i} f_i(t)\,dt.
\label{eq:lifetime_count}
\end{equation}

Defining the lifetime-averaged physiological frequency as

\begin{equation}
\bar f_i
\equiv
\frac{1}{L_i}
\int_{0}^{L_i} f_i(t)\,dt,
\label{eq:mean_frequency}
\end{equation}

one obtains the exact identity

\begin{equation}
N_i
=
\bar f_i L_i.
\label{eq:Nstar}
\end{equation}

The identity itself is elementary: frequency integrated over time produces
a cycle count. The empirical content of PBTE lies elsewhere. It is the
observation that, within an appropriately specified biological domain,
\(N_i\) does not vary in direct proportion to either lifespan or
physiological frequency. Instead, it concentrates around a characteristic
lifetime scale,

\begin{equation}
N_i
\simeq
N_{\star,C},
\label{eq:clade_invariant}
\end{equation}

where \(C\) denotes the relevant clade, physiological clock, and
normalization convention. Thus,

\begin{equation}
\bar f_i L_i
\simeq
N_{\star,C}.
\label{eq:PBTE_basic}
\end{equation}

Equation~\eqref{eq:PBTE_basic} is the elementary form of PBTE. Organisms
with rapid intrinsic clocks tend to traverse the lifetime cycle budget over
short chronological intervals, whereas organisms with slower clocks
distribute a comparable number of cycles over longer intervals. For the
cardiac clock of many non-primate placental mammals, the characteristic
scale is of order

\begin{equation}
N_{\star,C}
\sim
10^{9}
\quad
\text{cardiac cycles}.
\label{eq:cardiac_scale}
\end{equation}

PBTE does not require that every living system possess precisely the same
numerical value of \(N_\star\), nor that the cardiac clock be the appropriate
clock for every form of life. Rather, it proposes that lifetime biological
duration is constrained in the space of accumulated intrinsic events.
Different clades, physiological architectures, and environmental regimes
may require different effective clocks or explicit correction factors. The
near-billion-cycle cardiac scale is therefore a well-defined empirical
regularity within a particular biological domain, not an unrestricted
numerical constant applying without qualification to all living systems
\citep{Levine1997,SchmidtNielsen1984,TayeCardiac2026}.

The mouse--elephant comparison illustrates the physical meaning of this
relation. Representative values are

\begin{equation}
\bar f_{\rm mouse}
\simeq
600~\mathrm{min}^{-1},
\qquad
L_{\rm mouse}
\simeq
3~\mathrm{yr},
\end{equation}

and

\begin{equation}
\bar f_{\rm elephant}
\simeq
30~\mathrm{min}^{-1},
\qquad
L_{\rm elephant}
\simeq
70~\mathrm{yr}.
\end{equation}

Using

\begin{equation}
1~\mathrm{yr}
=
525{,}960~\mathrm{min},
\end{equation}

the corresponding lifetime cardiac counts are approximately

\begin{align}
N_{\rm mouse}
&\simeq
600
\times
3
\times
525{,}960
\nonumber\\
&\simeq
9.47\times10^{8},
\label{eq:N_mouse}
\\[6pt]
N_{\rm elephant}
&\simeq
30
\times
70
\times
525{,}960
\nonumber\\
&\simeq
1.10\times10^{9}.
\label{eq:N_elephant}
\end{align}

The mouse heart therefore operates approximately twenty times faster than
the elephant heart,

\begin{equation}
\frac{\bar f_{\rm mouse}}
     {\bar f_{\rm elephant}}
\simeq
20,
\end{equation}

whereas the elephant lives for approximately twenty-three times longer,

\begin{equation}
\frac{L_{\rm elephant}}
     {L_{\rm mouse}}
\simeq
23.
\end{equation}

The two ratios approximately compensate. In chronological units the lives
of the two organisms are radically unequal; in accumulated cardiac cycles
they are of the same order. The mouse traverses its biological trajectory
rapidly, while the elephant traverses a comparable intrinsic interval more
slowly. PBTE interprets this compensation as evidence that physiological
history is more naturally represented by accumulated internal activity than
by chronological duration alone.

For two organisms described relative to the same external time parameter,

\begin{equation}
\frac{d\theta_i}{d\theta_j}
=
\frac{f_i(t)}{f_j(t)}.
\label{eq:biological_transformation}
\end{equation}

This is a multiplicative transformation between biological rates. It is
not a Lorentz transformation. It contains no invariant limiting speed, no
relativity of simultaneity, and no spacetime metric. The relation expresses
only that two biological systems may accumulate intrinsic events at
different rates while sharing the same environmental time coordinate.
Figure~\ref{fig:worldlines} presents this chronological PBTE picture:
different organisms traverse comparable lifetime cycle budgets at markedly
different calendar rates.

The thermodynamic formulation developed in our previous work gives the
cycle count a physical interpretation. A living organism is an open
nonequilibrium system that maintains local organization by consuming free
energy and producing entropy. Let \(\Sigma_{p,i}\) denote cumulative
irreversible entropy production and let

\begin{equation}
\dot{\Sigma}_{p,i}
\equiv
\frac{d\Sigma_{p,i}}{dt}
\end{equation}

denote the corresponding entropy-production rate. The entropy produced per
physiological cycle is

\begin{equation}
\sigma_i
\equiv
\frac{d\Sigma_{p,i}}{d\theta_i}
=
\frac{\dot{\Sigma}_{p,i}}{f_i}.
\label{eq:sigma_cycle}
\end{equation}

Because whole-organism entropy production scales strongly with body mass,
cross-species comparisons are naturally expressed in mass-specific form.
For an organism of mass \(M_i\), define

\begin{equation}
\sigma_i^{\ast}
\equiv
\frac{1}{M_i}
\frac{d\Sigma_{p,i}}{d\theta_i}
=
\frac{\dot{\Sigma}_{p,i}}
     {M_i f_i}.
\label{eq:sigma_star_definition}
\end{equation}

If \(P_i\) denotes metabolic power and \(T_i\) the characteristic
temperature at which dissipation occurs, the leading thermodynamic estimate
is

\begin{equation}
\dot{\Sigma}_{p,i}
\simeq
\frac{P_i}{T_i}.
\label{eq:entropy_power}
\end{equation}

Consequently,

\begin{equation}
\sigma_i^{\ast}
\simeq
\frac{P_i/M_i}
     {T_i f_i}.
\label{eq:sigma_star}
\end{equation}

The differential thermodynamic closure of PBTE can therefore be written as

\begin{equation}
\frac{d\Sigma_{p,i}}{M_i}
=
\sigma_i^{\ast}\,d\theta_i.
\label{eq:entropy_cycle_relation}
\end{equation}

Using \(d\theta_i=f_i(t)\,dt\),

\begin{equation}
\frac{d\Sigma_{p,i}}{M_i}
=
\sigma_i^{\ast}(t)f_i(t)\,dt.
\label{eq:entropy_chronological}
\end{equation}

Integration over the lifetime gives

\begin{equation}
\frac{\Sigma_{p,i}^{\rm life}}{M_i}
=
\int_{0}^{N_i}
\sigma_i^{\ast}(\theta)\,d\theta.
\label{eq:lifetime_entropy_exact}
\end{equation}

If the cycle-averaged mass-specific entropy cost is defined by

\begin{equation}
\bar{\sigma}_i^{\ast}
\equiv
\frac{1}{N_i}
\int_{0}^{N_i}
\sigma_i^{\ast}(\theta)\,d\theta,
\label{eq:mean_sigma}
\end{equation}

then

\begin{equation}
\frac{\Sigma_{p,i}^{\rm life}}{M_i}
=
\bar{\sigma}_i^{\ast}N_i.
\label{eq:lifetime_entropy}
\end{equation}

Under the previously stated PBTE closure

\begin{equation}
\bar{\sigma}_i^{\ast}
\simeq
\sigma_{0,C}^{\ast},
\end{equation}

together with the empirical lifetime relation
\(N_i\simeq N_{\star,C}\), one obtains

\begin{equation}
\frac{\Sigma_{p,i}^{\rm life}}{M_i}
\simeq
\sigma_{0,C}^{\ast}N_{\star,C}.
\label{eq:thermodynamic_invariant}
\end{equation}

The cycle invariant and the thermodynamic invariant are therefore two
descriptions of the same lifetime constraint. The former counts biological
events; the latter weights those events by their irreversible entropy cost
\citep{TayeAging2026,TayeThermo2026}. Here
\(\Sigma_{p,i}\) denotes accumulated entropy production rather than entropy
stored within the organism. Living systems maintain internal organization
by exporting entropy, and PBTE concerns the irreversible throughput
associated with that maintenance.

An entropy-normalized biological age may consequently be defined as

\begin{equation}
A_{{\rm PBTE},i}(t)
=
\frac{1}{N_{\star,\rm ref}}
\int_{0}^{t}
\frac{\sigma_i^{\ast}(s)}
     {\sigma_{0,\rm ref}^{\ast}}
f_i(s)\,ds.
\label{eq:Apbte_chrono}
\end{equation}

Introducing the reference mass-specific lifetime entropy budget

\begin{equation}
\Sigma_{\rm ref}^{\ast}
\equiv
\sigma_{0,\rm ref}^{\ast}
N_{\star,\rm ref},
\label{eq:entropy_reference}
\end{equation}

Eq.~\eqref{eq:Apbte_chrono} is equivalently

\begin{equation}
A_{{\rm PBTE},i}(t)
=
\frac{\Sigma_{p,i}(t)/M_i}
     {\Sigma_{\rm ref}^{\ast}}.
\label{eq:Apbte_entropy}
\end{equation}

The physiological and aging consequences of this functional were developed
in the preceding PBTE work and are not re-derived here. The point relevant
to the present paper is that Eqs.~\eqref{eq:theta_definition} and
\eqref{eq:Apbte_chrono} have so far been parametrized by an external time
coordinate \(t\).

For ordinary terrestrial comparisons this convention is adequate because
the difference between terrestrial coordinate time and local proper time is
negligible at the level of biological measurement. It is not, however, the
fundamental formulation. Special relativity replaces the Newtonian notion
of a universal elapsed time with proper time attached to an individual
worldline. For an object moving at speed \(v_i\) relative to an inertial
coordinate system,

\begin{equation}
d\tau_i
=
dt
\sqrt{1-\frac{v_i^{2}}{c^{2}}}
=
\frac{dt}{\gamma_i},
\qquad
\gamma_i
=
\frac{1}
{\sqrt{1-v_i^{2}/c^{2}}}.
\label{eq:srt}
\end{equation}

The elapsed proper time along the moving worldline is therefore smaller
than the elapsed coordinate time assigned by the external inertial frame.
This does not mean that the traveler experiences a local slowing of
physiology. In the traveler's instantaneous rest frame, atomic transitions,
chemical reactions, metabolism, cardiac activity, and cognition proceed
normally. Time dilation arises when elapsed durations along distinct
worldlines are compared.

The physical status of Eq.~\eqref{eq:srt} is independent of PBTE. The
relation is supported by measurements of relativistic particle lifetimes,
transported atomic clocks, precision optical clocks, and satellite timing
systems
\citep{Bailey1977,HafeleKeating1972a,HafeleKeating1972b,Chou2010,Ashby2003}.
These observations are not presented as experimental confirmations of
biological time equivalence. Their relevance is more fundamental: they
establish that local physical processes are parametrized by proper time.
Muon decay, atomic oscillation, chemical change, and physiological activity
do not obey separate relativistic laws.

The same principle admits a geometric expression in a general spacetime.
For a timelike worldline \(x_i^\mu\), with metric signature
\((- , + , + , +)\),

\begin{equation}
d\tau_i
=
\frac{1}{c}
\sqrt{
-g_{\mu\nu}\,
dx_i^\mu dx_i^\nu
}.
\label{eq:gendtau}
\end{equation}

Equation~\eqref{eq:srt} is the flat-spacetime inertial limit of
Eq.~\eqref{eq:gendtau}. The special-relativistic case is sufficient to
expose the central PBTE issue: if physiology evolves locally according to
proper time, then the expression

\begin{equation}
d\theta_i
=
f_i(t)\,dt
\end{equation}

is only a nonrelativistic shorthand. Its invariant replacement is

\begin{equation}
d\theta_i
=
f_i(\tau_i)\,d\tau_i.
\label{eq:theta_proper}
\end{equation}

Correspondingly, the thermodynamic increment becomes

\begin{equation}
\frac{d\Sigma_{p,i}}{M_i}
=
\sigma_i^{\ast}(\tau_i)
f_i(\tau_i)\,d\tau_i.
\label{eq:entropy_proper}
\end{equation}

Relative to an external coordinate time \(t\),

\begin{equation}
\frac{d\theta_i}{dt}
=
f_i(\tau_i)
\frac{d\tau_i}{dt}.
\label{eq:theta_coordinate_rate}
\end{equation}

For uniform motion in flat spacetime this reduces to

\begin{equation}
\frac{d\theta_i}{dt}
=
\frac{f_i(\tau_i)}{\gamma_i}.
\label{eq:theta_sr_rate}
\end{equation}

This relation establishes the division of physical roles on which the
remainder of the paper is based. Relativity determines the amount of proper
time accumulated along an organism's worldline. PBTE determines the amount
of physiological progression and irreversible thermodynamic expenditure
accumulated per unit proper time. Relativistic biological time dilation is
therefore not produced by a separate slowing of metabolism. It is the
consequence of expressing an intrinsically proper-time-parametrized
biological process relative to an external coordinate clock.

\section{Covariant Relativistic PBTE and Its Physical Consequences}
\label{sec:relpbte}

The preceding formulation identifies two intrinsic quantities associated
with a living system. Relativistic proper time, \(\tau\), measures the
invariant duration accumulated along the system's spacetime worldline,
whereas biological proper time, \(\theta\), measures the accumulated number
of physiologically meaningful events occurring during that duration. The
relativistic completion of PBTE follows by requiring every biological and
thermodynamic rate to be defined locally with respect to \(\tau\), rather
than with respect to an observer-dependent coordinate time \(t\).

Let \(\Sigma_p\) denote the cumulative irreversible entropy production of
the organism and let

\begin{equation}
\dot{\Sigma}_p(\tau)
\equiv
\frac{d\Sigma_p}{d\tau}
\label{eq:proper_entropy_rate}
\end{equation}

denote the entropy-production rate measured in its local rest frame. The
entropy-normalized PBTE age is then

\begin{equation}
\boxed{
A_{\rm PBTE}(\tau)
=
\frac{1}{\Sigma_{\rm ref}}
\int_{\tau_0}^{\tau}
\dot{\Sigma}_p(\tau')\,d\tau'
}
\label{eq:central}
\end{equation}

where \(\Sigma_{\rm ref}\) is the specified reference lifetime entropy
budget. Equation~\eqref{eq:central} is the central result of the present
work. It states that biological age is an irreversible thermodynamic
functional evaluated along the organism's worldline. The value of
\(A_{\rm PBTE}\) is therefore independent of the coordinate system used to
describe that worldline.

The corresponding aging rate relative to an external coordinate time
follows directly from the chain rule:

\begin{equation}
\frac{dA_{\rm PBTE}}{dt}
=
\frac{1}{\Sigma_{\rm ref}}
\frac{d\Sigma_p}{d\tau}
\frac{d\tau}{dt}.
\label{eq:chain}
\end{equation}

For uniform inertial motion in flat spacetime,

\begin{equation}
\frac{d\tau}{dt}
=
\sqrt{1-\frac{v^2}{c^2}}
=
\frac{1}{\gamma},
\qquad
\gamma
=
\frac{1}{\sqrt{1-v^2/c^2}},
\end{equation}

and Eq.~\eqref{eq:chain} becomes

\begin{equation}
\boxed{
\frac{dA_{\rm PBTE}}{dt}
=
\frac{\dot{\Sigma}_p}
{\gamma\Sigma_{\rm ref}}
}
\label{eq:srpbte}
\end{equation}

with \(\dot{\Sigma}_p=d\Sigma_p/d\tau\) evaluated in the organism's local
rest frame. In the weak-field and slow-motion limit,

\begin{equation}
\frac{dA_{\rm PBTE}}{dt}
\simeq
\frac{\dot{\Sigma}_p}{\Sigma_{\rm ref}}
\left(
1+\frac{\Phi}{c^2}
-\frac{v^2}{2c^2}
\right),
\label{eq:grpbte}
\end{equation}

where \(\Phi\) is the Newtonian gravitational potential along the
worldline. Equations~\eqref{eq:srpbte} and \eqref{eq:grpbte} separate two
physically distinct contributions to externally observed biological
aging. The local entropy-production rate describes the physiological and
thermodynamic state of the organism, whereas \(d\tau/dt\) describes the
geometry of its worldline. Neither contribution is reducible to the other.

This distinction resolves a common causal misconception. A moving
organism does not age more slowly because velocity directly suppresses its
metabolism. In its instantaneous local rest frame, metabolism, cardiac
activity, molecular turnover, neural activity, and every other internal
process proceed normally. The reduction observed from another frame arises
because less proper time accumulates along the moving worldline between the
chosen comparison events. The local relation

\begin{equation}
\frac{dA_{\rm PBTE}}{d\tau}
=
\frac{\dot{\Sigma}_p}{\Sigma_{\rm ref}}
\label{eq:local_aging_rate}
\end{equation}

is unchanged by uniform inertial motion. What changes is the conversion
between local proper time and external coordinate time.

The same reasoning applies to the physiological cycle count. The
nonrelativistic expression \(d\theta=f(t)\,dt\) is replaced by

\begin{equation}
d\theta
=
f(\tau)\,d\tau,
\label{eq:theta_proper}
\end{equation}

where \(f(\tau)\) is the local physiological frequency measured per unit
proper time. Relative to an external coordinate time,

\begin{equation}
\frac{d\theta}{dt}
=
f(\tau)\frac{d\tau}{dt},
\label{eq:theta_external}
\end{equation}

and for inertial motion,

\begin{equation}
\frac{d\theta}{dt}
=
\frac{f(\tau)}{\gamma}.
\label{eq:freqdt}
\end{equation}

Every local biological clock is therefore assigned the same global
relativistic factor when expressed relative to the same external
coordinate time. Heart rate, respiration, circadian phase, cell division,
chemical turnover, and molecular aging are not independently dilated.
They are locally normal processes sharing a common proper-time
parameterization.

The distinction between PBTE scaling and relativistic time dilation can
now be stated precisely. The mouse--elephant comparison concerns two
organisms with different intrinsic physiological frequencies occupying
approximately the same spacetime environment. Their difference is
primarily biological:

\begin{equation}
\frac{d\theta_{\rm mouse}}
     {d\theta_{\rm elephant}}
\simeq
\frac{f_{\rm mouse}}
     {f_{\rm elephant}}
\simeq
20.
\end{equation}

By contrast, the relativistic comparison concerns systems following
different spacetime worldlines. Their difference is geometric and is
expressed through the ratio of their proper-time rates. For two organisms
\(i\) and \(j\) described in a common coordinate system, both effects
combine as

\begin{equation}
\frac{d\theta_i/dt}{d\theta_j/dt}
=
\underbrace{
\frac{f_i(\tau_i)}
     {f_j(\tau_j)}
}_{\text{physiological factor}}
\,
\underbrace{
\frac{d\tau_i/dt}
     {d\tau_j/dt}
}_{\text{spacetime factor}}.
\label{eq:combined}
\end{equation}

Equation~\eqref{eq:combined} provides a compact unification without
confusing the two mechanisms. If the organisms occupy the same terrestrial
environment, the proper-time ratio is effectively unity and the relation
reduces to ordinary PBTE scaling. If identical organisms follow different
worldlines while maintaining the same local physiology, the physiological
ratio is unity and only the relativistic factor remains. If both physiology
and worldline differ, their effects compose multiplicatively.

A simple example makes the interpretation transparent. Consider an
organism traveling at \(v=0.8c\) relative to Earth. The Lorentz factor is

\begin{equation}
\gamma
=
\frac{1}{\sqrt{1-0.8^2}}
=
\frac{5}{3},
\qquad
\frac{d\tau}{dt}
=
0.6.
\end{equation}

During one year of Earth coordinate time, the traveler accumulates
\(0.6\) years of proper time. If its locally measured resting heart rate is
\(70~\mathrm{min}^{-1}\), its accumulated cardiac count is

\begin{equation}
\Delta\theta_{\rm traveler}
=
70
\times
0.6
\times
525{,}960
\simeq
2.21\times10^7
\end{equation}

beats. An otherwise identical organism remaining at rest relative to Earth
accumulates

\begin{equation}
\Delta\theta_{\rm Earth}
=
70
\times
525{,}960
\simeq
3.68\times10^7
\end{equation}

beats during the same Earth-coordinate interval. The ratio is

\begin{equation}
\frac{\Delta\theta_{\rm traveler}}
     {\Delta\theta_{\rm Earth}}
=
0.6
=
\frac{1}{\gamma}.
\end{equation}

The traveler nevertheless measures a normal local heart rate of
\(70~\mathrm{min}^{-1}\). After correcting for signal-propagation and
Doppler effects, Earth assigns a rate of \(42\) beats per Earth-coordinate
minute. The reduced accumulated heartbeat count is thus a consequence of
the smaller proper-time interval, not of impaired cardiac function.

Gravity produces the corresponding effect through gravitational
time dilation. For a stationary worldline at radius \(r\) in the
Schwarzschild geometry,

\begin{equation}
d\tau
=
dt
\sqrt{
1-\frac{2GM}{rc^2}
}.
\label{eq:schwarzschild}
\end{equation}

An organism deeper in the gravitational potential accumulates less proper
time per unit of the distant observer's coordinate time. If its local
physiological state remains unchanged, its externally assigned PBTE aging
rate is reduced by the same factor. Near Earth's surface, two stationary
systems separated vertically by a small height \(h\) have the approximate
fractional clock-rate difference

\begin{equation}
\frac{\Delta(d\tau/dt)}
     {d\tau/dt}
\simeq
\frac{gh}{c^2}.
\label{eq:height_shift}
\end{equation}

For \(h=1~\mathrm{m}\), this quantity is approximately
\(1.1\times10^{-16}\). The effect is measurable with modern optical clocks
but overwhelmingly smaller than ordinary biological variability. Its
importance here is therefore conceptual rather than biomedical: even at
laboratory scales, the fundamental temporal variable remains proper time.

The proper-time formulation may also be written covariantly. Let
\(s^\mu\) denote the entropy four-current of the biological system. The
local second law is expressed as

\begin{equation}
\nabla_\mu s^\mu
=
\sigma
\geq
0,
\label{eq:entropy_current}
\end{equation}

where \(\sigma\) is the local entropy-production density. For an organism
whose spatial extent is small relative to the relevant curvature scale, a
representative center-of-mass worldline and an instantaneous local
rest-frame volume \(V(\tau)\) may be used. The total entropy-production
rate is then

\begin{equation}
\dot{\Sigma}_p(\tau)
=
\int_{V(\tau)}
\sigma(x,\tau)\,
dV_{\rm proper}.
\label{eq:total_entropy_rate}
\end{equation}

Substitution into Eq.~\eqref{eq:central} gives

\begin{equation}
\boxed{
A_{\rm PBTE}
=
\frac{1}{\Sigma_{\rm ref}}
\int_{\tau_0}^{\tau_1}
\int_{V(\tau)}
\nabla_\mu s^\mu\,
dV_{\rm proper}\,d\tau
}
\label{eq:covariant}
\end{equation}

which is independent of the coordinates chosen to describe the
worldtube. Equation~\eqref{eq:covariant} is the manifestly covariant form
of relativistic PBTE. It reduces to the ordinary terrestrial expression
when spacetime curvature is weak, velocities are small, and
\(d\tau\simeq dt\).

The geometrical content required for this construction is correspondingly
modest. Biological proper time may be represented by the one-dimensional
line element

\begin{equation}
d\ell_{\rm B}
=
f(\tau)\,d\tau
=
d\theta,
\label{eq:biological_line_element}
\end{equation}

so that the accumulated biological path length is

\begin{equation}
\ell_{\rm B}
=
\int f(\tau)\,d\tau.
\end{equation}

This representation is useful because it distinguishes the amount of
proper time supplied by spacetime from the number of intrinsic events
accumulated during that interval. It does not define an additional
spacetime metric, introduce Lorentz symmetry into biology, or imply an
invariant biological speed analogous to \(c\). The relativistic metric is
fundamental; the biological line element is a state-dependent measure
constructed on top of it.

The framework yields several clear physical consequences. Uniform
inertial motion produces no locally detectable physiological change.
Identical local biological clocks following different worldlines acquire
the same relativistic dilation factor when compared in a common external
coordinate system. Physiological perturbations and relativistic effects
remain separable:

\begin{equation}
\frac{dA_{\rm PBTE}}{dt}
=
\left(
\frac{\dot{\Sigma}_p^{\rm local}}
     {\Sigma_{\rm ref}}
\right)
\left(
\frac{d\tau}{dt}
\right).
\label{eq:factorization}
\end{equation}

Radiation, microgravity, inflammation, circadian disruption, stress, or
disease may alter the first factor, while velocity and gravitational
potential alter the second. Real spaceflight can therefore accelerate
local physiological deterioration even while relativistic motion reduces
the amount of proper time accumulated per Earth year. The net biological
outcome is determined by their product, not by either contribution alone.

Equation~\eqref{eq:factorization} also states the principal empirical
consistency condition of the theory. After local environmental and
physiological effects are accounted for, independent biological clocks
should remain mutually synchronized in proper time and should acquire the
same global relativistic factor relative to an external coordinate time.
A violation in which different isolated internal processes acquired
different velocity-dependent factors would contradict the proper-time
formulation. Direct biological tests at relativistically significant
velocities are not currently feasible, and terrestrial gravitational
effects are far below normal biological noise. The experimentally
established particle and atomic-clock evidence therefore validates the
proper-time factor used here, but it does not by itself validate the
biological invariant, the entropy-per-cycle closure, or the PBTE aging
functional. Those remain independent biological hypotheses requiring
their own comparative and physiological tests.

\begin{figure}[t]
\centering
\includegraphics[width=0.82\textwidth]{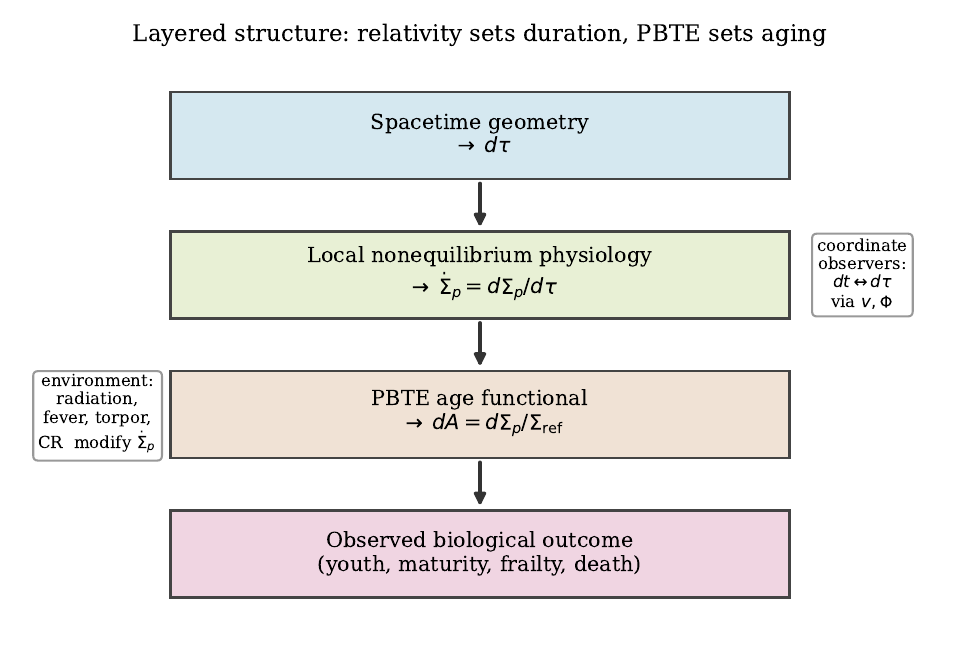}
\caption{Logical structure of relativistic PBTE. Spacetime geometry
determines the proper-time increment \(d\tau\). Local physiological state
determines the intrinsic biological frequency \(f(\tau)\) and entropy
production rate \(\dot{\Sigma}_p(\tau)\). Biological proper time and PBTE
age are then accumulated as \(d\theta=f\,d\tau\) and
\(dA_{\rm PBTE}=\dot{\Sigma}_p\,d\tau/\Sigma_{\rm ref}\). Velocity and
gravitational potential modify the mapping between coordinate time and
proper time; environmental and pathological conditions modify the local
physiological terms.}
\label{fig:hierarchy}
\end{figure}

\section{Deviation From Temporal Balance: Localised Symmetry Breaking and Applications}
\label{sec:deviation}
The relativistic functional of Section~\ref{sec:relpbte} describes an organism that traverses
its biological history along a single worldline. Sections~\ref{sec:basics}--\ref{sec:relpbte}
treated the normal case, in which an organism completes life having consumed an
approximately clade-invariant intrinsic budget. We now ask a complementary question that the
proper-time formulation makes precise: how should one describe systems that depart from that
balance, and how is the description modified when the relevant rates are referred to proper
time rather than coordinate time? The constructions below are advanced as testable PBTE
hypotheses rather than established results; aging itself, and its damage dynamics, are treated
in the companion work \citep{TayeAging2026} and are not re-examined here. Our focus is the
parametrisation of deviation along the worldline and two applications---inflammatory bias and
viral latency---that are natural consequences of it.

\subsection{A temporal order parameter referred to proper time}
To quantify how much of an organism's intrinsic budget has been spent, it is convenient to
introduce a single dimensionless coordinate. Let \(N(\tau)=\int_{\tau_0}^{\tau}f(\tau')\,d\tau'\)
be the intrinsic cycle count accumulated along the worldline, with \(f\) the local
physiological frequency per unit proper time of Eq.~\eqref{eq:theta_proper}. Define the
\emph{temporal order parameter}
\begin{equation}
\zeta(\tau)=\frac{N(\tau)}{N_\star}
=\frac{1}{N_\star}\int_{\tau_0}^{\tau}f(\tau')\,d\tau',
\label{eq:zeta}
\end{equation}
in which \(N_\star\) is the reference intrinsic budget introduced in
Section~\ref{sec:basics}. The quantity \(\zeta\) is a biological odometer: \(\zeta=0\) at the
start of the trajectory and, on the reference manifold, \(\zeta=1\) at its end. Because the
integral in Eq.~\eqref{eq:zeta} is taken over proper time, \(\zeta\) is itself an invariant of
the worldline, independent of the external coordinate used to describe it, exactly as
\(A_{\rm PBTE}\) was shown to be in Eq.~\eqref{eq:central}. Using the thermodynamic closure of
Section~\ref{sec:basics}, \(d\Sigma_p=\sigma_0\,d\theta\), and the lifetime budget
\(\Sigma_\star=\sigma_0N_\star\), Eq.~\eqref{eq:zeta} acquires an equivalent thermodynamic form,
\begin{equation}
\zeta(\tau)=\frac{\Sigma_p(\tau)}{\Sigma_\star},
\label{eq:zeta_entropy}
\end{equation}
so that \(\zeta\) is simultaneously the fraction of the intrinsic cycle budget and the fraction
of the lifetime entropy budget consumed. Comparison with Eq.~\eqref{eq:central} shows that
\(\zeta\) and the PBTE age \(A_{\rm PBTE}\) coincide when the reference budgets are chosen
consistently; \(\zeta\) is therefore the natural state variable in which deviations from
temporal balance are expressed. Three regimes follow directly from Eq.~\eqref{eq:zeta}: a
balanced trajectory with \(\zeta\to1\) as the worldline ends, a \emph{hypertemporal} trajectory
with \(\zeta>1\) in which the budget is over-consumed, and a \emph{hypotemporal} trajectory with
\(\zeta<1\) in which it is under-consumed.

\subsection{Relaxation dynamics, physiological noise, and inflammatory bias}
A healthy organism does not merely possess a value of \(\zeta\); it actively maintains \(\zeta\)
near its reference trajectory against perturbation. To represent this regulation, define the
fractional deviation \(\phi(\tau)=\zeta(\tau)-\tau/\tau_{\rm exp}\) from the expected reference
path, where \(\tau_{\rm exp}\) is the proper time at which a balanced organism reaches
\(\zeta=1\). Near the reference manifold the simplest dynamics consistent with homeostatic
restoration are linear, giving an Ornstein--Uhlenbeck relaxation along the worldline,
\begin{equation}
\frac{d\phi}{d\tau}=-\frac{\phi(\tau)}{\tau_\zeta}+\eta(\tau)+u(\tau),
\label{eq:OU}
\end{equation}
in which each term has a definite role. The restoring term \(-\phi/\tau_\zeta\) pulls the system
back toward balance, and the \emph{temporal resilience time} \(\tau_\zeta>0\) sets the rate of
recovery: small \(\tau_\zeta\) denotes a resilient system that corrects perturbations quickly,
large \(\tau_\zeta\) a fragile one. The term \(\eta(\tau)\) represents zero-mean stochastic
physiological fluctuations, \(\langle\eta(\tau)\rangle=0\),
\(\langle\eta(\tau)\eta(\tau')\rangle=2D_\eta\,\delta(\tau-\tau')\), with noise intensity
\(D_\eta\). The term \(u(\tau)\) is a sustained deterministic bias---an external drive that
displaces the organism systematically from balance. Chronic inflammation is the paradigmatic
positive bias: by raising metabolic demand and repair load it acts in Eq.~\eqref{eq:OU} as a
persistent \(u>0\) that pushes \(\phi\) upward, driving the system toward the hypertemporal
regime; a sustained suppression of metabolic rate enters instead as \(u<0\). The advantage of
expressing Eq.~\eqref{eq:OU} in proper time is that the resilience time \(\tau_\zeta\) and the
bias \(u\) are then intrinsic properties of the organism, cleanly separable from the relativistic
factor \(d\tau/dt\) that governs how the worldline is observed externally.

\begin{figure}[t]
\centering
\includegraphics[width=0.66\textwidth]{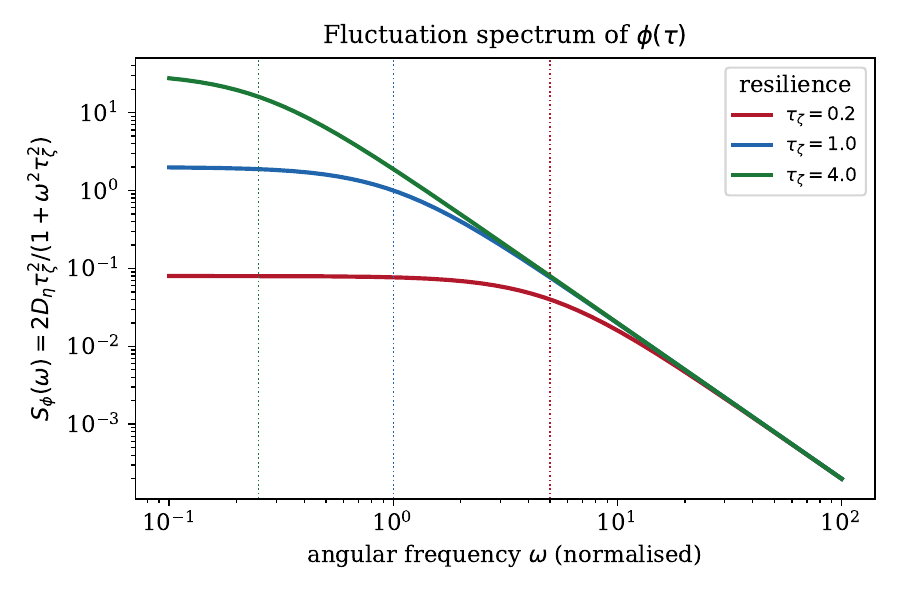}
\caption{Fluctuation spectrum of the temporal-deviation variable \(\phi\). The Lorentzian
spectrum of Eq.~\eqref{eq:spectrum} is shown for three values of the resilience time
\(\tau_\zeta\); dotted lines mark the corner frequencies \(\omega_c=1/\tau_\zeta\). As
\(\tau_\zeta\) grows, spectral weight migrates to low frequencies, so that a fragile system
recovering slowly from perturbation displays slower, more correlated fluctuations. The
prediction that such low-frequency amplification precedes a transition from temporal balance to
the hypertemporal regime is directly testable in physiological time series.}
\label{fig:spectrum_zeta}
\end{figure}

Equation~\eqref{eq:OU} makes a falsifiable prediction. Its stationary fluctuation spectrum is
the Lorentzian
\begin{equation}
S_\phi(\omega)=\frac{2D_\eta\,\tau_\zeta^{2}}{1+\omega^{2}\tau_\zeta^{2}},
\label{eq:spectrum}
\end{equation}
where \(\omega\) is the angular frequency conjugate to proper time. As the resilience time
\(\tau_\zeta\) increases, the corner frequency \(\omega_c=1/\tau_\zeta\) of
Eq.~\eqref{eq:spectrum} shifts to lower values and fluctuations become slower and more strongly
correlated. Figure~\ref{fig:spectrum_zeta} plots Eq.~\eqref{eq:spectrum} for three resilience
times and shows this leftward migration of spectral weight. A loss of temporal resilience is
therefore expected to announce itself, before any change in mean physiological state, as a
slowing and low-frequency amplification of fluctuations in measurable signals such as heart-rate
variability, temperature, glucose, sleep rhythm, and inflammatory markers---a biological
analogue of critical slowing down near a transition.

\subsection{Localised symmetry breaking: a spatially resolved order parameter}
\begin{figure}[t]
\centering
\includegraphics[width=0.96\textwidth]{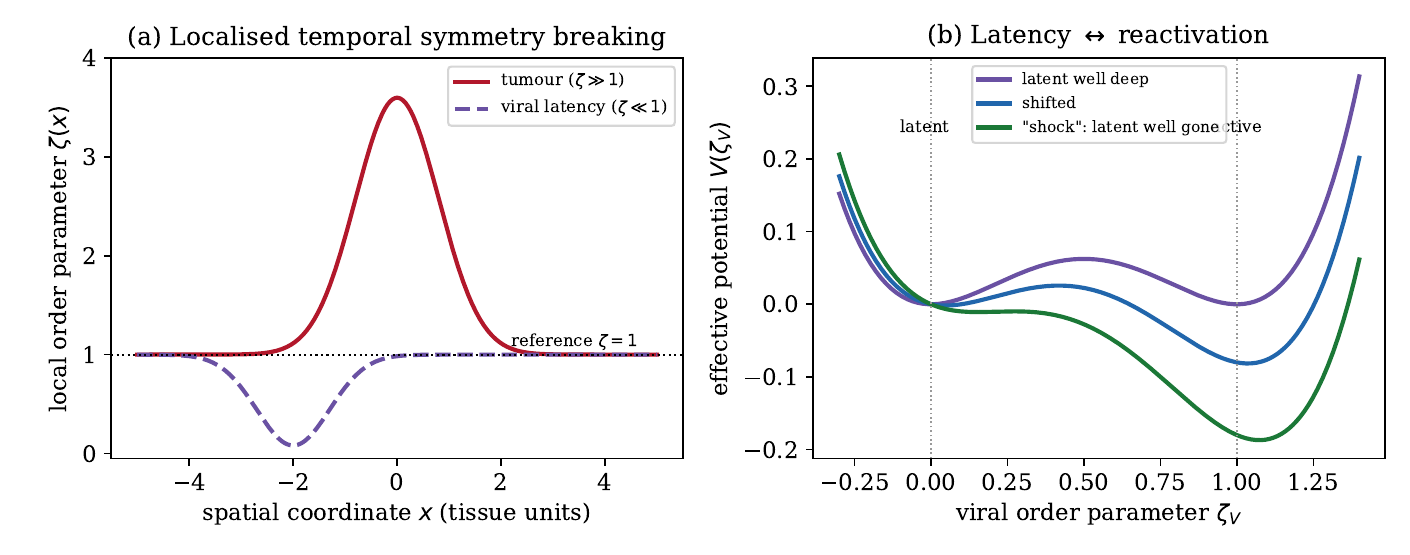}
\caption{Localised temporal symmetry breaking. (a) Stationary spatial profiles of the local
order parameter \(\zeta(x)\) from Eq.~\eqref{eq:reactdiff}: a domain of accelerated local time
(\(\zeta\gg1\), solid) and a domain of near-arrested local time (\(\zeta\ll1\), dashed),
relaxing toward the homeostatic value \(\zeta=1\) over the healing length
\(\ell_\zeta=\sqrt{D_\zeta\tau_\zeta}\). (b) Effective bistable potential
\(V(\zeta_V)\) for an embedded pathogen: a deep latent well near \(\zeta_V\approx0\) and an
active-replication state near \(\zeta_V\approx1\) are separated by a barrier; a latency-reversal
drive shifts the potential so that the latent well disappears.}
\label{fig:localzeta}
\end{figure}

In a multicellular organism the whole-organism worldline may remain balanced while a localised
tissue or an embedded pathogen runs an internal clock that departs sharply from the
organism-wide average. To represent this, promote the order parameter to a field over the
proper-time slices of the worldtube,
\begin{equation}
\zeta(x,\tau)=\frac{f(x,\tau)\,\tau_{\rm loc}(x,\tau)}{N_\star},
\label{eq:localzeta}
\end{equation}
where \(f(x,\tau)\) is the local intrinsic frequency and \(\tau_{\rm loc}\) the local
persistence time at proper-time-labelled position \(x\). Homeostatic coupling between
neighbouring tissue, together with a local pathological drive, gives the reaction--diffusion
dynamics
\begin{equation}
\tau_\zeta\,\partial_\tau\zeta=-(\zeta-1)+D_\zeta\nabla^{2}\zeta+S(x,\tau),
\label{eq:reactdiff}
\end{equation}
whose three terms are, in order, relaxation of the local clock toward the homeostatic value
\(\zeta=1\), spatial coupling that synchronises adjacent tissue with effective diffusivity
\(D_\zeta\), and a source \(S(x,\tau)\) encoding persistent local drivers of temporal
dysregulation. The balance of the first two terms defines a characteristic healing length
\(\ell_\zeta=\sqrt{D_\zeta\tau_\zeta}\) over which a local defect in biological time is repaired
by its surroundings. Equation~\eqref{eq:reactdiff} reduces to the whole-organism relaxation
\eqref{eq:OU}, without noise, when \(\zeta\) is spatially uniform and \(S\) plays the role of
the bias \(u\). Figure~\ref{fig:localzeta}(a) shows two opposite stationary profiles of
Eq.~\eqref{eq:reactdiff} that organise the applications below.

\subsection{Application: localised acceleration and viral latency}
Two limiting forms of Eq.~\eqref{eq:reactdiff} are of direct biological interest, and both are
forms of spontaneous temporal symmetry breaking. In the first, a localised population sustains a
source \(S>0\) strong enough to hold \(\zeta(x,\tau)\gg1\) within a bounded region while the
surrounding tissue remains near \(\zeta=1\). The accompanying elevation of local entropy
production per cell---of which the preferential use of aerobic glycolysis is a recognised
signature \citep{Vander2009}---and the steep spatial gradients \(\nabla\zeta\) at the domain
boundary disrupt the intercellular temporal coordination that normally maintains tissue
homeostasis. Proliferative pathology is, in this language, a domain of locally accelerated
biological time embedded in a balanced organism; we state this as a PBTE interpretation to be
tested, not as a mechanistic claim about causation.

In the second limit the local frequency is driven nearly to zero. During the latent phase of an
infection the replication frequency \(f_V\approx0\), so that, from Eq.~\eqref{eq:localzeta},
\begin{equation}
\zeta_V^{(\ell)}=\frac{f_V\,\tau_{\rm loc}}{N_\star}\ll1,
\label{eq:latency}
\end{equation}
signalling near-arrest of the pathogen's biological time while the host's proper time continues
to advance normally. Latency is thus a state in which two coupled systems sharing the same
spacetime worldtube occupy opposite extremes of the order parameter. Reactivation is the escape
of \(\zeta_V\) from the latent value toward \(\zeta_V\approx1\), naturally modelled as a
transition in the bistable effective potential of Figure~\ref{fig:localzeta}(b): a deep well near
\(\zeta_V\approx0\) (latency) and a second minimum near \(\zeta_V\approx1\) (active replication)
separated by a barrier. Latency-reversal strategies correspond to deforming that potential until
the latent well becomes unstable, driving the deliberate transition
\(\zeta_V\ll1\to\zeta_V\approx1\) \citep{Deeks2012}. The same construction places host--pathogen
encounters in the language of interacting clocks: productive infection requires the pathogen and
the host response to occupy commensurate values of \(\zeta\) within a finite window, a condition
expressed compactly by the proper-time-referred frequencies of Section~\ref{sec:relpbte}.

\subsection{Application: restoration as optimal control}
If deviation from temporal balance is undesirable, its correction can be posed as an optimal
control problem, which is the practical payoff of expressing biological state through \(\zeta\).
One seeks the intervention \(u(\tau)\) in Eq.~\eqref{eq:OU} that minimises the quadratic cost
\begin{equation}
J[u]=\int_{\tau_0}^{\tau_1}\!\Bigl[\phi(\tau)^2+\lambda\,u(\tau)^2\Bigr]\,d\tau,
\label{eq:control}
\end{equation}
in which the first term penalises residual deviation of the biological clock from its reference
trajectory and the second penalises the burden of intervention, with \(\lambda>0\) setting their
relative weight. The minimiser of Eq.~\eqref{eq:control} is a linear state feedback,
\(u^\star(\tau)\propto-\phi(\tau)\): one measures how far the system has drifted and applies the
smallest correction that restores balance. Candidate physiological channels through which such a
control could act include circadian entrainment by timed light and feeding, intermittent caloric
restriction, mild thermal modulation, and, for the latency application, latency-reversal agents
that reshape the potential of Figure~\ref{fig:localzeta}(b). Because \(\zeta\) is estimable in
principle from wearable physiological data, Eqs.~\eqref{eq:OU} and \eqref{eq:control} together
suggest a closed-loop architecture---continuous estimation of \(\zeta\), comparison with the
reference trajectory, and delivery of a timed intervention---in which relativistic corrections
enter only through the conversion \(d\tau/dt\) of Section~\ref{sec:relpbte} and are negligible
for terrestrial application. These proposals define an empirical programme; their clinical
status remains to be established.

\section{Discussion and Conclusion}
\label{sec:discussion}

The purpose of the present work is not to construct a biological
alternative to special or general relativity. It is to place PBTE within
the temporal structure already established by relativity. This distinction
prevents two opposite conceptual errors. The first is to regard biological
proper time as merely metaphorical. In PBTE, biological proper time is an
operational quantity: it is the accumulated count of specified intrinsic
events,

\begin{equation}
\theta
=
\int f(\tau)\,d\tau.
\end{equation}

The second error is to identify that quantity with Einsteinian proper time.
The two are not identical. Relativistic proper time is a geometric
duration determined by the spacetime metric; biological proper time is a
state-dependent accumulation of physiological events occurring during
that duration. The correct relation is hierarchical:

\begin{equation}
t
\longrightarrow
\tau
\longrightarrow
\theta
\longrightarrow
A_{\rm PBTE}.
\label{eq:temporal_hierarchy}
\end{equation}

Coordinate time \(t\) describes events relative to a chosen reference
system. Proper time \(\tau\) gives the invariant duration along the
organism's worldline. Biological proper time \(\theta\) counts internal
physiological events accumulated along that worldline. PBTE age
\(A_{\rm PBTE}\) weights irreversible physiological activity relative to a
reference lifetime budget.

The principal theoretical advance is therefore a covariant completion of
the PBTE functional. In its nonrelativistic form, biological age was
expressed as an integral over terrestrial chronological time. That
expression is adequate whenever \(d\tau\simeq dt\), as it is for ordinary
comparative biology. It is not, however, invariant under changes of
spacetime trajectory. Equation~\eqref{eq:central} removes this dependence
by defining biological age directly along the organism's proper-time
worldline.

This result clarifies the relationship between the familiar
mouse--elephant comparison and Einsteinian time dilation. A mouse and an
elephant accumulate biological events at different rates because their
local physiological frequencies differ. Identical twins following
different relativistic worldlines accumulate different amounts of
biological history because their elapsed proper times differ. The first is
a physiological rescaling; the second is a geometric effect. They meet in
Eq.~\eqref{eq:combined}, which shows that biological and relativistic
clock-rate differences compose multiplicatively while remaining
conceptually and experimentally distinguishable.

The formulation also places an important limit on the interpretation of
entropy within PBTE. Entropy production does not generate the spacetime
metric, and it is not presented here as the cause of relativistic time
dilation. Atomic clocks and unstable particles accumulate proper time
without possessing biological entropy budgets. Conversely, living systems
are distinctive not because they alone possess time, but because they
transform proper-time duration into organized physiological history,
memory, repair, adaptation, damage, and eventually loss of function.
Spacetime supplies duration; irreversible thermodynamics supplies
direction; biological organization converts irreversible activity into
age.

The phrase ``time is entropy's arrow; life is entropy's clock'' is
therefore defensible only when interpreted with this hierarchy in mind.
The thermodynamic arrow distinguishes irreversible histories, while the
PBTE clock records a biologically selected subset of entropy-producing
events. A nonliving object may undergo entropy production and physical
change without possessing a regulated lifetime cycle budget. A living
organism, by contrast, maintains itself through continuous energetic
throughput and thereby converts irreversible expenditure into an
internally structured trajectory.

Nature more generally contains a hierarchy of characteristic internal
times. A star possesses a nuclear-burning timescale, a mountain a
geological weathering timescale, a radioactive nucleus a decay timescale,
a cell a division-and-repair timescale, and a virus a host-dependent
replication timescale. These are not separate spacetime dimensions but
process coordinates generated by the characteristic rates of the systems
concerned. PBTE identifies a more specific structure within living matter.
Living systems do not merely undergo change; they actively maintain a
low-entropy organization through continuous free-energy consumption,
repair, molecular turnover, regulation, and entropy export. Their internal
time is therefore simultaneously kinematic and thermodynamic: physiological
cycles supply the ticks, while entropy production supplies the irreversible
cost of advancing them. A multicellular organism moreover contains many
such clocks at once---cardiac, respiratory, circadian, immune, neural,
metabolic, reproductive, and molecular---which need not remain perfectly
synchronized. PBTE does not require that one clock be universally dominant;
it provides a framework in which these clocks may be coarse-grained,
compared, and weighted according to their contribution to organismal
maintenance and irreversible biological burden.

Several claims lie outside the present theory. The framework does not
claim that only living systems measure time, that a body causes
relativistic time dilation through metabolic slowing, that zero entropy
production implies the cessation of proper time, or that the approximate
lifetime invariant \(N_\star\) has the exact status of a fundamental
constant. Relativistic proper time is fundamental within the spacetime
theory; PBTE is an emergent biological model whose invariant, closure
relations, and reference budgets must be evaluated empirically within
specified domains.

For ordinary terrestrial organisms, relativistic corrections are
negligible compared with physiological variation. Differences caused by
sleep, infection, temperature, stress, diet, activity, inflammation, and
genetic background exceed the kinematic and gravitational factors by many
orders of magnitude. The covariant formulation is nevertheless necessary
at the conceptual level. A proposed law of biological age should identify
the correct independent temporal variable even when the correction is too
small to matter in everyday practice.

This interpretation has concrete consequences, because it decomposes
biological aging into three experimentally distinguishable elements: the
rate of internal ticking, the irreversible cost associated with each tick,
and the effective lifetime budget over which those costs can be sustained.
Torpor and hibernation primarily lower the local accumulation rate of
physiological cycles; improved mitochondrial coupling may reduce entropy
production per effective cycle; enhanced repair and damage clearance may
change the fraction of dissipation that contributes to persistent aging;
and evolutionary adaptation may alter the effective budget itself. These
mechanisms should not be treated as interchangeable merely because each can
extend chronological lifespan, since each generates a different PBTE
signature and therefore different empirical predictions. The same
decomposition allows disease to be read as temporal dysregulation: chronic
inflammation, cancer, sustained sympathetic activation, metabolic disease,
or repeated repair failure may create a hypertemporal state by raising
physiological rate, raising irreversible cost per cycle, or accelerating the
conversion of energetic throughput into persistent damage. A clinical
assessment would then require both a state measure---how much biological
distance has already been traversed---and a rate measure---how rapidly the
remaining trajectory is being consumed. The framework extends naturally to
spaceflight, where environmental stressors modify local physiology while
velocity and gravitational potential independently modify the available
proper time, the two contributions remaining separable through the
factorized relativistic PBTE law of Eq.~\eqref{eq:factorization}.

The wider implication is a change in the status assigned to calendar time
throughout biology. Chronological time remains indispensable as an external
coordinate, but it is no longer assumed to be a complete measure of
development, aging, disease progression, or ecological interaction. Equal
calendar intervals can contain unequal quantities of physiological work,
entropy production, repair, replication, and accumulated biological history.
Comparative experiments may therefore become more informative when organisms
are matched not only by chronological age but also by physiological-cycle
count or entropy-normalized PBTE age. Host--pathogen dynamics can be viewed
as interactions between clocks running at different intrinsic rates, in which
immune success or failure may depend on whether those rates synchronize
within a finite response window; ecological relationships similarly couple
organisms whose reproductive, metabolic, and seasonal clocks occupy
different parts of a temporal hierarchy. At the conceptual level, PBTE
removes the presumption of a single universal biological present. Spacetime
supplies proper duration, thermodynamics supplies irreversible direction,
physiology supplies intrinsic ticks, and PBTE supplies the transformation
and normalization through which different living times become comparable.
What changes is not the physics of time but the biological meaning assigned
to its passage: the same external hour can contain radically different
amounts of life.

The resulting synthesis can be stated compactly. Relativity determines
how much proper time is available along a worldline. Local physiology
determines how rapidly biological cycles and entropy production accumulate
during that proper time. PBTE converts the resulting irreversible
expenditure into a normalized aging coordinate:

\begin{equation}
\boxed{
A_{\rm PBTE}
=
\frac{1}{\Sigma_{\rm ref}}
\int_{\tau_0}^{\tau_1}
\dot{\Sigma}_p(\tau)\,d\tau
}
\end{equation}

with the coordinate-time representation

\begin{equation}
\boxed{
\frac{dA_{\rm PBTE}}{dt}
=
\frac{\dot{\Sigma}_p}{\Sigma_{\rm ref}}
\frac{d\tau}{dt}.
}
\end{equation}

The organism does not make relativistic time run slowly. It undergoes
physiological change during the proper time accumulated along its
worldline. Relativity determines the duration of that worldline; PBTE
describes the irreversible biological history written within it.

\section*{Declarations}
\addcontentsline{toc}{section}{Declarations}

\noindent
\textbf{Data availability.}
No new empirical dataset is analyzed in this theoretical manuscript.
Figures are generated from the analytic relations defined in the text.

\medskip

\noindent
\textbf{Competing interests.}
The author declares no competing interests.

\medskip

\noindent
\textbf{Funding.}
No external funding is declared for this manuscript.

\medskip

\noindent
\textbf{Ethics statement.}
No new research involving human participants or animals was conducted.
\appendix
\section{Supplementary Information}
\label{app:supplement}

This Supplementary Information collects the graphical illustrations,
algebraic details, numerical checks, and falsifiability conditions directly
required by the relativistic formulation of PBTE. It is intentionally
restricted to material that supports the central result of the main text:
biological proper time and entropy-normalized biological age must be
integrated with respect to relativistic proper time along the organism's
worldline.

\begin{figure}[htbp]
\centering
\includegraphics[width=0.80\textwidth]{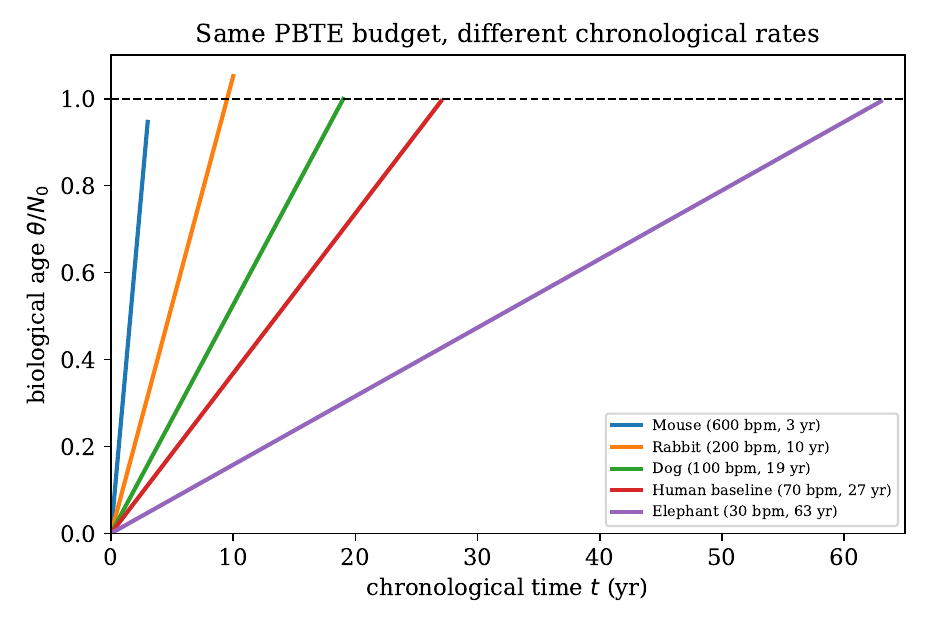}
\caption{PBTE trajectories for representative organisms in a common
terrestrial coordinate time. The ordinate is the normalized accumulated
cycle count \(\theta(t)/N_0\). High-frequency organisms traverse the
reference cycle budget rapidly, whereas low-frequency organisms distribute
a comparable intrinsic path length over a longer chronological interval.
The dashed horizontal line denotes \(\theta/N_0=1\). These trajectories
represent physiological rate differences and not relativistic time
dilation.}
\label{fig:worldlines}
\end{figure}

Figure~\ref{fig:worldlines} illustrates the ordinary nonrelativistic PBTE
picture. Organisms with different intrinsic physiological frequencies
accumulate biological proper time at different rates when described relative
to the same terrestrial coordinate time. For an approximately constant
cardiac frequency \(f_H\),

\begin{equation}
\theta(t)
=
f_H\,t\,C_{\rm yr},
\qquad
C_{\rm yr}
=
525{,}960~\mathrm{min\,yr^{-1}},
\label{eq:supp_theta}
\end{equation}

when \(f_H\) is expressed in beats per minute and \(t\) in years. Normalizing
by the reference lifetime count \(N_0=10^9\) gives the fraction of the
reference cardiac-cycle budget traversed by chronological age \(t\).

\begin{figure}[htbp]
\centering
\includegraphics[width=0.72\textwidth]{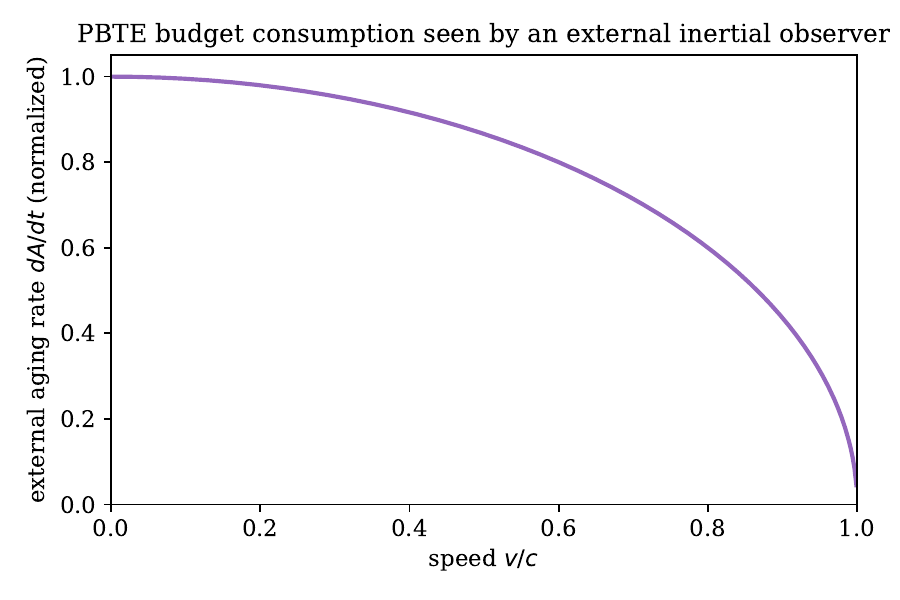}
\caption{Externally assigned PBTE aging rate for an organism with fixed
local entropy production per unit proper time. The plotted quantity is the
rate relative to the same organism at rest,
\((dA/dt)_v/(dA/dt)_0=1/\gamma\). The traveler experiences normal local
physiology; the reduction arises because less proper time accumulates along
the moving worldline during a fixed interval of external coordinate time.}
\label{fig:pbterate}
\end{figure}

Figure~\ref{fig:pbterate} shows the distinct relativistic effect. Consider
an organism whose local entropy-production rate per unit proper time remains
fixed. Relative to an inertial coordinate time \(t\), its PBTE aging rate is

\begin{equation}
\frac{dA_{\rm PBTE}}{dt}
=
\frac{\dot{\Sigma}_p}{\Sigma_{\rm ref}}
\frac{d\tau}{dt},
\label{eq:supp_aging_rate}
\end{equation}

where \(\dot{\Sigma}_p=d\Sigma_p/d\tau\). For uniform motion in flat
spacetime,

\begin{equation}
\frac{d\tau}{dt}
=
\sqrt{1-\frac{v^2}{c^2}}
=
\frac{1}{\gamma},
\label{eq:supp_lorentz_factor}
\end{equation}

and therefore

\begin{equation}
\frac{dA_{\rm PBTE}}{dt}
=
\frac{\dot{\Sigma}_p}
{\gamma\Sigma_{\rm ref}}.
\label{eq:supp_sr_rate}
\end{equation}

The decrease shown in Figure~\ref{fig:pbterate} is not a local suppression
of metabolism. It is the reduction in proper time accumulated per unit
external coordinate time.

A third illustration is useful only to preserve the distinction between
physiological and relativistic clock-rate changes. In torpor or hibernation,
the organism alters its local physiological frequency. This is a genuine
biological slowing of cycle accumulation per unit local proper time. It is
therefore conceptually different from Lorentz dilation, for which the local
frequency remains unchanged and only the relation between \(d\tau\) and
\(dt\) changes.

\begin{figure}[htbp]
\centering
\includegraphics[width=0.78\textwidth]{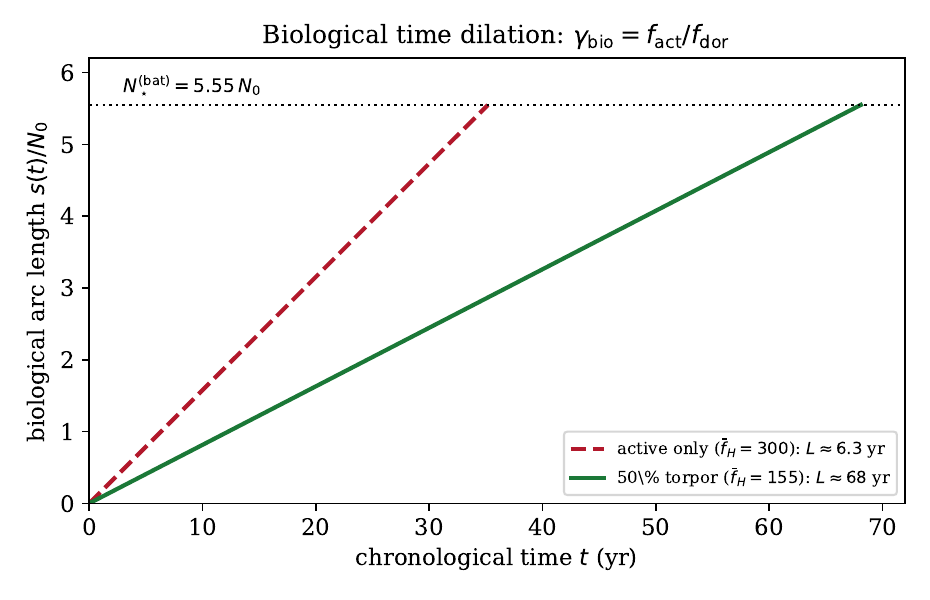}
\caption{Physiological extension of chronological lifespan in a two-state
active--torpor model. The dormant state lowers the locally measured
physiological frequency and therefore reduces \(d\theta/d\tau\). The same
cycle budget may consequently be distributed over a longer chronological
interval. This mechanism is biological rather than relativistic: torpor
changes the local clock rate \(f(\tau)\), whereas relativistic dilation
changes \(d\tau/dt\) while leaving local physiology unchanged.}
\label{fig:biodilation}
\end{figure}

The combined transformation law follows directly from the separation of
these two effects. Let organism \(i\) follow the timelike worldline
\(x_i^\mu\). Its relativistic proper-time increment is

\begin{equation}
d\tau_i
=
\frac{1}{c}
\sqrt{
-g_{\mu\nu}\,
dx_i^\mu dx_i^\nu
},
\label{eq:supp_proper_time}
\end{equation}

for metric signature \((- , + , + , +)\). Biological proper time is
accumulated locally according to

\begin{equation}
d\theta_i
=
f_i(\tau_i)\,d\tau_i.
\label{eq:supp_biological_time}
\end{equation}

If the worldline is parametrized by a coordinate time \(t\), then

\begin{equation}
\frac{d\theta_i}{dt}
=
f_i(\tau_i)
\frac{d\tau_i}{dt}.
\label{eq:supp_coordinate_rate}
\end{equation}

For two organisms \(i\) and \(j\) described in the same coordinate system,

\begin{equation}
\boxed{
\frac{d\theta_i/dt}
     {d\theta_j/dt}
=
\frac{f_i(\tau_i)}
     {f_j(\tau_j)}
\,
\frac{d\tau_i/dt}
     {d\tau_j/dt}
}
\label{eq:supp_combined}
\end{equation}

follows without additional assumptions. The first ratio is physiological;
the second is geometric. If the organisms occupy the same terrestrial
environment, then \(d\tau_i/dt\simeq d\tau_j/dt\), and
Eq.~\eqref{eq:supp_combined} reduces to the ordinary PBTE clock-rate
relation. If identical organisms maintain the same local physiology while
following different worldlines, then \(f_i=f_j\), and the expression reduces
to the relativistic proper-time ratio. When physiology and worldline both
differ, the two factors compose multiplicatively.

The same derivation applies to entropy-normalized biological age. For

\begin{equation}
A_i
=
\frac{1}{\Sigma_{\rm ref}}
\int
\dot{\Sigma}_{p,i}(\tau_i)\,d\tau_i,
\label{eq:supp_age_functional}
\end{equation}

one obtains

\begin{equation}
\frac{dA_i}{dt}
=
\frac{\dot{\Sigma}_{p,i}(\tau_i)}
     {\Sigma_{\rm ref}}
\frac{d\tau_i}{dt}.
\label{eq:supp_age_coordinate}
\end{equation}

Consequently, the externally assigned aging-rate ratio of two organisms is

\begin{equation}
\frac{dA_i/dt}{dA_j/dt}
=
\frac{\dot{\Sigma}_{p,i}(\tau_i)}
     {\dot{\Sigma}_{p,j}(\tau_j)}
\,
\frac{d\tau_i/dt}
     {d\tau_j/dt},
\label{eq:supp_age_ratio}
\end{equation}

provided the same reference normalization is used. This factorization is
the formal basis for separating environmental and physiological influences
from relativistic effects.

Table~\ref{tab:supp_sr} gives the special-relativistic proper-time factor
for representative velocities. The final column may be read either as the
proper time accumulated during one unit of external coordinate time or as
the external PBTE aging rate relative to an identical organism at rest,
assuming equal local entropy production.

\begin{table}[htbp]
\centering
\caption{Special-relativistic proper-time and PBTE rate factors. The
normalized rate assumes that \(dA/d\tau\) is identical at all velocities.}
\label{tab:supp_sr}
\begin{tabular}{@{}cccc@{}}
\toprule
\(v/c\) &
\(\gamma\) &
\(d\tau/dt=1/\gamma\) &
\((dA/dt)_v/(dA/dt)_0\) \\
\midrule
0.10 & 1.005 & 0.995 & 0.995 \\
0.50 & 1.155 & 0.866 & 0.866 \\
0.80 & 1.667 & 0.600 & 0.600 \\
0.90 & 2.294 & 0.436 & 0.436 \\
0.99 & 7.089 & 0.141 & 0.141 \\
\bottomrule
\end{tabular}
\end{table}

At \(v=0.8c\), for example, \(d\tau/dt=0.6\). During one Earth year of
uniform inertial motion, the traveler accumulates \(0.6\) years of proper
time. A locally measured cardiac frequency of
\(70~\mathrm{min}^{-1}\) therefore produces

\begin{equation}
\Delta\theta_{\rm traveler}
=
70
\times
0.6
\times
525{,}960
\simeq
2.21\times10^7
\end{equation}

cardiac cycles. An otherwise identical organism at rest relative to Earth
accumulates

\begin{equation}
\Delta\theta_{\rm Earth}
=
70
\times
525{,}960
\simeq
3.68\times10^7
\end{equation}

cycles over the same Earth-coordinate interval. Their ratio is
\(0.6=1/\gamma\). The traveler nevertheless measures a normal local
frequency of \(70~\mathrm{min}^{-1}\).

The relativistic extension also imposes explicit consistency and
falsifiability conditions. First, the biological foundation remains
independently testable: within a prespecified domain, the lifetime quantity

\begin{equation}
N_i
=
\int_0^{L_i} f_i(t)\,dt
\end{equation}

must exhibit substantially weaker systematic variation than either
frequency or lifespan separately. A strong residual dependence on body
mass, temperature, or phylogeny after the stated PBTE corrections would
falsify the proposed invariant for that domain.

Second, the thermodynamic closure requires the independently estimated
entropy cost per effective physiological cycle,

\begin{equation}
\sigma_i^\ast
=
\frac{1}{M_i}
\frac{d\Sigma_{p,i}}{d\theta_i},
\end{equation}

to remain sufficiently concentrated within the population or clade for
which approximate constancy is claimed. Systematic variation over several
orders of magnitude would invalidate the simple closure and would require
a different physiological weighting.

Third, the relativistic component predicts no locally detectable effect of
uniform inertial motion. An organism enclosed in a freely moving laboratory
must not infer its uniform velocity from local metabolism, cardiac rhythm,
chemical reaction rates, or any other internal experiment. A model
predicting direct velocity-dependent physiological suppression in the local
rest frame would violate the principle of relativity.

Fourth, after local physiological perturbations are controlled, all
internal clocks must inherit the same global proper-time factor relative to
a common external coordinate time. Cardiac cycles, respiration, molecular
turnover, circadian phase, and other local processes may possess different
intrinsic frequencies, but relativity cannot assign them different Lorentz
factors. Any such process-specific kinematic dilation would contradict the
proper-time formulation.

Fifth, environmental effects must remain separate from spacetime effects.
Radiation, microgravity, altered exercise, confinement, circadian
disruption, stress, inflammation, and disease may change the local
entropy-production rate \(\dot{\Sigma}_p\). Velocity and gravitational
potential change \(d\tau/dt\). The observable aging rate is their product,

\begin{equation}
\frac{dA_{\rm PBTE}}{dt}
=
\left(
\frac{\dot{\Sigma}_p^{\rm local}}
     {\Sigma_{\rm ref}}
\right)
\left(
\frac{d\tau}{dt}
\right).
\label{eq:supp_factorization}
\end{equation}

Real spaceflight may therefore increase local physiological burden while
relativistic motion decreases the amount of proper time accumulated per
Earth year. Those effects are physically distinct and may, in principle,
oppose one another.

Finally, the present framework does not claim that PBTE has independently
confirmed special or general relativity. Particle-decay measurements,
atomic clocks, optical clocks, and satellite timing establish the
proper-time structure adopted here. They do not establish the PBTE
lifetime invariant, the entropy-per-cycle closure, or the biological aging
functional. Conversely, evidence for PBTE would not constitute a new test
of Lorentz invariance unless a biological clock were compared under
conditions in which relativistic effects could be isolated from local
physiological perturbations. The contribution of the present theory is the
consistent composition of these independently testable structures.


\end{document}